\documentclass[10pt,aps,prl,twocolumn,preprint,nofootinbib]{revtex4}

\usepackage{amsmath}    
\usepackage{mathtools}
\usepackage{graphicx}   
\usepackage{color}
\usepackage{verbatim}

\usepackage{cancel}

\begin{document}

\title{Size Effects in the Ginzburg-Landau Theory}
\author{Miguel C N Fiolhais}
\email{miguel.fiolhais@cern.ch}   
\affiliation{Department of Physics, City College of the City University of New York, 160 Convent Avenue, New York 10031, NY, USA}
\affiliation{LIP, Department of Physics, University of Coimbra, 3004-516 Coimbra, Portugal}
\author{Joseph L Birman}
\email{birman@sci.ccny.cuny.edu}   
\affiliation{Department of Physics, City College of the City University of New York, 160 Convent Avenue, New York 10031, NY, USA}
\date{\today}
\begin{abstract} 
The Ginzburg-Landau theory is analyzed in the case of small dimension superconductors, a couple of orders of magnitude above the coherence length, where the theory is still valid but quantum fluctuations become significant. In this regime, the potential around the expectation value is approximated to a quadratic behavior, and the ground-state derived from the Klein-Gordon solutions of the Higgs-like field. The ground-state energy is directly compared to the condensation energy, and used to extract new limits on the size of superconductors at zero Kelvin and near the critical temperature. \\

\end{abstract}

\maketitle

\section{Introduction}

The Ginzburg-Landau (GL) theory~\cite{ginzburg}, proposed in 1950, is perhaps the most successful macroscopic description of superconductivity, proceeding the famous phenomenological description of the electromagnetic field in a superconductor by the London brothers~\cite{london}. The proposed macroscopic quantum theory by Ginzburg and Landau makes use of a quartic potential and the Higgs mechanism of spontaneous symmetry breaking~\cite{anderson,higgs,englert,guralnik} to generate a local mass term of the vector potential. As such, it successfully describes the Meissner-Ochsenfeld effect~\cite{meissner,fiolhais,essen,fiolhais2}, for the magnetic field flux expulsion, the phenomenology associated with the phase transition~\cite{halperin,kleinert,kleinert2,kleinert3}, and also predicts the existence of a coherence length in superconductors, resulting from the local scalar order field fluctuations. Furthermore, in 1957, Abrikosov \cite{abrikosov} predicted the penetration of strong magnetic fields in type-II superconductors through quantum vortices, giving farther credibility to the GL model. The GL-theory can be regarded nowadays as the three-dimensional version of the 3+1-dimensional scalar quantum electrodynamics, studied in detail by Coleman and Weinberg \cite{coleman,weinberg}, in the early seventies, as an attempt to generate the scalar field mass through radiative corrections.


The size limitations and effects in superconductivity have been under study for several decades~\cite{parmenter,janocko,guo,zgirski,clark}. In this letter, a new method is developed to derive the approximate size limit of type-II superconductors in the Ginzburg-Landau theory. The application of the Ginzburg-Landau theory is, first and foremost, constrained by the magnitude of the coherence length, below which such a macroscopic theory is no longer valid. However, for type-II superconductors, with small coherence lengths, the quantum fluctuations may become significant enough to impose a new physical limit to superconductivity, above the coherence length. In addition to this, the Ginzburg-Landau theory, in the London approximation, shall only be applied to type-II superconductors with large GL parameters, also known as ``clean'' superconductors, as opposed to type-I superconductors, yielding non-local effects, where the Pippard's model must be taken into account~\cite{pippard}. The presence of a macroscopic massive scalar Higgs-like field pervading the superconducting region in the Ginzburg-Landau theory, corresponding to the collective excitation of the Cooper pairs in the lattice~\cite{fiolhais3,fiolhais4}, implies the existence of a ground-state energy resulting from the quantum fluctuations of the scalar field between the superconductor walls. While for macroscopic superconductors, these fluctuations are usually negligible as the scalar field behaves continuously in the classical approximation, they become relevant for small sized superconductors, leading to a discretization of energy levels, and therefore, of the supercurrents as well. In particular, for small superconductor, the ground state energy is expected to increase, eventually to the point where it surpasses the condensation energy and restores the vacuum symmetry of the GL potential. These quantum fluctuations can be parameterized by a quadratic approximation around the minimum of the potential, leading to the Klein-Gordon equation for the scalar field. Therefore, the ground-state and the allowed energy levels can be directly extracted from the Klein-Gordon solutions for the scalar field in a box. Such approximation is, of course, limited to small fluctuations, but may also be extrapolated to large fluctuations, in order to provide an interesting lower limit on the size of superconductors.

The Ginzburg-Landau ground-state energies, derived from the Klein-Gordon solutions, and the corresponding size limits predictions at absolute zero and near the critical temperature, are presented in this letter. In particular, as the condensation energy vanishes near the second-order phase transition, the lower size limit increases, possibly reaching macroscopic dimensions.



\section{The Ginzburg-Landau Theory and the Higgs mechanism}

At zero kelvin, the hamiltonian of the Ginzburg-Landau theory of superconductivity can be written as:
\begin{eqnarray}
 \mathcal{H} (\psi, {\nabla} \psi, \mathbf{A}, {\nabla} \mathbf{A}) & = &  \frac{1}{2m_{\textrm{e}}} | \left ( - i \hbar {\nabla} - 2 e \mathbf{A} \right ) \psi |^2 \nonumber \\
&+& \alpha |\psi|^2 + \frac{\beta}{2} |\psi|^4  \nonumber \\
&+& \frac{1}{2\mu_0} \left ( {\nabla} \times \mathbf{A} \right )^2 \, ,
\label{hamiltonian}
\end{eqnarray}
with the order parameter $\psi(x) = \rho (x) \textrm{e}^{i\theta (x)}$, where $\rho(x)$ and $\theta (x)$ are real fields, $2e$ is the electric charge of the Cooper pairs, and the real constants $\alpha$ and $\beta$ 
give the strength of the quadratic and quartic terms, respectively. The vector field is represented by $\mathbf{A}$. The ground state of the potential, $V(\psi) = \alpha |\psi|^2 + \frac{\beta}{2} |\psi|^4 $, is particularly interesting for a negative mass parameter $\alpha$, 
\begin{equation}
\langle \psi \rangle^2 = \rho_0^2 = - \frac{\alpha}{\beta} \, ,
\end{equation}
corresponding to an infinite number of degenerate states. After the spontaneous symmetry breaking, \emph{i.e.} fixing the gauge to $\theta (x) = 0$, the energy density becomes
\begin{eqnarray}
 \mathcal{H} (\psi, {\nabla} \psi, \mathbf{A}, {\nabla} \mathbf{A}) & = &  \frac{\hbar^2}{2m_{\textrm{e}}} \left ( {\nabla} \rho\right )^2 + V (\rho) + \frac{2e^2\rho^2}{m_{\textrm{e}}} \mathbf{A}^2 \nonumber \\
&+& \frac{1}{2\mu_0} \left ( {\nabla} \times \mathbf{A} \right )^2 \, .
\label{hamiltonian2}
\end{eqnarray}
The mass term of the vector field, resulting from the spontaneous symmetry breaking, and also known as the Meissner-Higgs mass term~\cite{kleinert3}, suppresses the magnetic field inside, with a corresponding London penetration length,
\begin{equation}
\lambda_L = \sqrt{\frac{m_{\textrm{e}}}{4\mu_0 e^2 \rho_0^2}} \, . 
\label{eq:density}
\end{equation}
On the other hand, the scalar field mass term can be associated with the coherence length,
\begin{equation}
\xi = \sqrt{\frac{\hbar^2}{4m_{\textrm{e}}|\alpha|}} \, .
\end{equation}
Finally, the thermodynamic critical magnetic field can also be extracted from the condensation energy, \emph{i.e.} the necessary energy to restore the vacuum symmetry, which leads to
\begin{equation}
B_c = \frac{1}{4}\frac{\hbar}{e}\frac{1}{\lambda_L\xi} \, .
\end{equation}
At finite temperature near $T_c$, the quadratic parameter varies linearly with the temperature,
\begin{equation}
\alpha (T) \approx \alpha_0 \left ( 1 - \frac{T}{T_c} \right ) \, .
\end{equation}


\section{The Quadratic Approximation and the Klein-Gordon Equation Solutions}

\begin{figure}[!h]
\begin{center}
\includegraphics[height=6.cm]{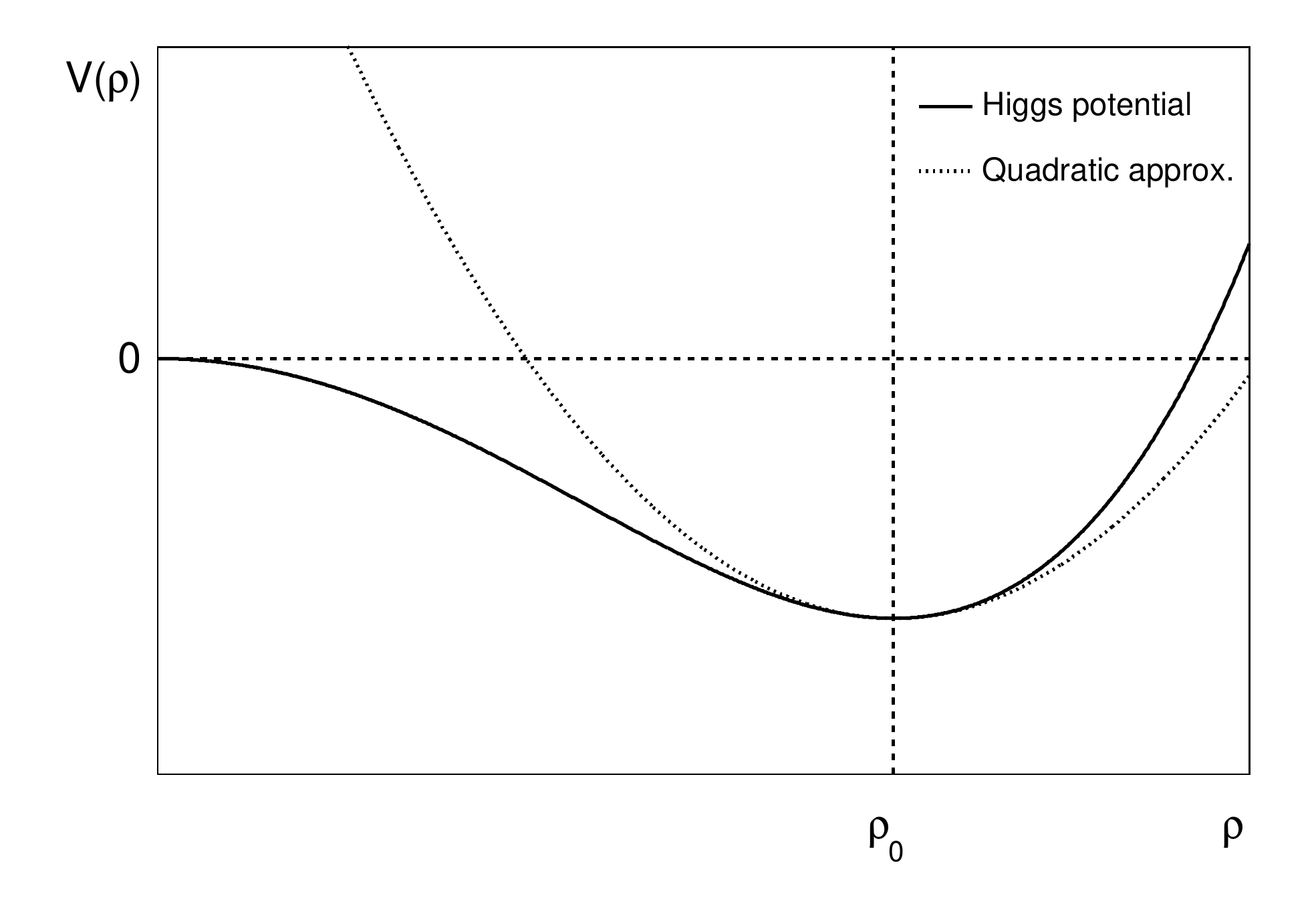}
\caption{Quadratic approximation (dashed line) of the Higgs potential (full line) around the expectation value.}
\label{size}
\end{center}
\end{figure}

As mentioned before, the quartic potential of the Ginzburg-Landau theory can be approximated to a parabola around the expectation value, by expanding it to the second order of Taylor series,
\begin{equation}
V(\rho) \approx  \, - \frac{\alpha^2}{2\beta} + 2\alpha \left (\rho -\rho_0 \right )^2 \, ,
\label{eq:london}
\end{equation}
where the fluctuations around the expectation value can be expressed in terms of a Higgs-like field, \mbox{$h(x) = \rho(x) - \rho_0$}. The quadratic approximation around the expectation value is represented in Figure \ref{size}.

In the particular case of small dimensions superconductors, where significant quantum fluctuations arise, the four-dimensional version of the Ginzburg-Landau theory, the Coleman-Weinberg model, comprising time dependencies, becomes a more accurate description. Assuming there is no external magnetic field, and the internal contributions to the electromagnetic field are too small\footnote{This is only valid if the superconducting region is considered to have a local net zero charge density, corresponding to the macroscopic limit, \emph{i.e.} the Ginzburg-Landau regime.}, the dynamics of the scalar Higgs field in the quadratic approximation of the Coleman-Weinberg model results in the following time-dependent hamiltonian,
\begin{equation}
 \mathcal{H} (h, {\nabla} h) = \frac{\hbar^2}{2m_{\textrm{e}}} \left ( \left ( {\nabla} h \right )^2 - \frac{1}{c^2}\left ( \partial_t h \right )^2 \right ) - \frac{\alpha^2}{2\beta} + 2\alpha h ^2  .
\label{hamiltonian3}
\end{equation}
As such, in this approximation, the dynamic equation for the scalar field simplifies to a Klein-Gordon equation,
\begin{equation}
\frac{1}{c^2}\frac{\partial^2 h(x)}{\partial t^2}  - \nabla^2 h (x) + \frac{2 m_{\textrm{e}} |\alpha|}{\hbar^2} h (x) = 0 \, .
\end{equation}
It should be stressed, however, that this approximation is more favorable for temperatures near criticality, rather than at zero Kelvin, as the expectation value of the Higgs potential approaches to zero giving rise to a second order phase transition. In any case, it provides a good estimate on the magnitude of the ground-state energy one shall expect in such potential, even at low temperature. 

The plane wave solutions for the Higgs-like field are, therefore, of the type,
\begin{equation}
h(\mathbf{r}, t) \sim e^{i(\mathbf{k}\cdot\mathbf{r}-\omega t)} \, ,
\end{equation}
where $-|\mathbf{k}|^2+\omega^2=\mu^2$. In the particular case the scalar field is constrained to a small sized three-dimensional box, like in a superconductor with small dimensions, the different energy levels are given by the solutions for the free particle of the infinite square potential~\cite{alberto},
\begin{equation}
E_{n_x,n_y,n_z} =   \sqrt{2m_{\textrm{e}}c^2|\alpha| + \frac{ \sum_i n_i^2 \pi^2 \hbar^2 c^2}{ L^2}}  \, .
\label{energylevels}
\end{equation}
In the particular case the box dimensions are much larger than the coherence length ($L >> \xi$), the second term in equation (\ref{energylevels}) becomes less relevant. Note that this condition is necessary to keep the validity of the Ginzburg-Landau theory intact in this regime. As a result, the kinetic component of the ground-state energy is approximately given by the non-relativist limit,
\begin{equation}
E_1^{\textrm{k}} \approx  \frac{3\sqrt{2}}{4} \frac{ \pi^2 \hbar^2 c}{\sqrt{m_{\textrm{e}}|\alpha|} L^2} =  \frac{3\sqrt{2}}{2} \frac{ \pi^2 \hbar \xi c}{ L^2} \, .
\label{groundstate}
\end{equation}

\section{The Ground State Energy as the Size Limit of Superconductivity}

In small superconductors, but still larger than the coherence length, as the ground-state energy arises, it may surpass the condensation energy and restore the vacuum symmetry. As such, one can establish a limit on the size of the superconductor by imposing that the energy density of its ground-state is equal to the condensation energy,
\begin{equation}
\frac{E_1}{L^3} = \frac{B_{\textrm{c}}^2}{2\mu_0} \, ,
\end{equation}
which leads to,
\begin{equation}
L^5 = 48\sqrt{2} \frac{e^2 \mu_0 c}{\hbar} \xi^3 \lambda^2 \, .
\end{equation}
or,
\begin{equation}
L = 1.44 \, \xi^{3/5} \lambda^{2/5} \, .
\end{equation}
At zero Kelvin, the size limit of a superconductor in the Ginzburg-Landau theory depends on both the coherence and penetration lengths, prominently on the first. 
For that reason, this result is only valid for clean superconductors with a Ginzburg-Landau parameter~\cite{fiolhaiskleinert} much larger than one, $k_{\textrm{GL}} >> 1$, 
so that the minimum allowed scale can be at least one order of magnitude larger than the coherence length, enabling the validity of the Ginzburg-Landau theory.

On the other hand, as the quadratic term vanishes near criticality, the ground-state energy of the scalar field at finite temperature in this regime is no longer in the non-relativistic limit,
\begin{equation}
E_1 \approx \frac{\sqrt{3} \pi \hbar c}{L} \, .
\end{equation} 
As a result, the minimal allowed scale in the Ginzburg-Landau theory near the critical temperature is,
\begin{equation}
L^4 (T) =  \frac{32\mu_0 e^2}{\sqrt{3}\pi\hbar} \lambda_L^2(T) \xi^2(T) \, ,
\end{equation}
or,
\begin{equation}
L(T) \approx 0.86 \sqrt{\lambda_L(T) \xi(T)} \, .
\end{equation}
It should be noted, as well, that since the minimum allowed scale near criticality also depends on both the coherence length and the London penetration depth, its temperature dependence near the critical temperature is expected to increase at a smaller rate than the coherence length. As such, the validity of this model is quite questionable for temperatures extremely close to the criticality, even in clean superconductors.

\section{Conclusions}

The lower size limits of clean high-temperature superconductors were derived in the framework of the Ginzburg-Landau theory at zero Kelvin and near the critical temperature. The existence of a ground-state for the scalar Higgs-like field was predicted and derived, and compared with the condensation energy for extremely small superconductors. As result, the minimum allowed scale was found to depend on both the coherence length and the London penetration depth in both regimes, within the range of applicability of this model, limiting it to clean high-temperature superconductors. As the experimental results on the size limits of clean high-temperature superconductors are, to the best of our knowledge, currently quite limited, we would like to challenge experimental groups to test the predicted results obtained with this model. A concurrent result would provide additional to the already successful Ginzburg-Landau theory, while the opposite would oblige further scrutiny on its range of validity.




\begin{thebibliography}{5}
\bibitem{ginzburg} V.L. Ginzburg and L.D. Landau, \emph{Zh. Eksp. Teor. Fiz.} \textbf{20} (1950) 1064.
\bibitem{london} F. London \and H. London, \emph{Proc. Roy. Soc.}, \textbf{A149} (866) (1935) 71.
\bibitem{anderson} P.W. Anderson, \emph{Phys. Rev.} \textbf{130} (1963) 439. 
\bibitem{englert} F. Englert and R. Brout, \emph{Phys. Rev. Lett.} \textbf{13} (1964) 321.
\bibitem{higgs} P.W. Higgs, \emph{Phys. Rev. Lett.} \textbf{13} (1964) 508. 
\bibitem{guralnik} G. S. Guralnik, C.R. Hagen, and T.W.B. Kibble, \emph{Phys. Rev. Lett.} \textbf{13} (1964) 585.
\bibitem{meissner} W. Meissner and R. Ochsenfeld, \emph{Naturwiss} \textbf{21} (1933) 787.
\bibitem{fiolhais} M.C.N. Fiolhais, H. Ess\'en, C. Provid\^encia and A. Nordmark, \emph{Prog. Electromagn. Res. B (USA)} \textbf{27} (2011) 187. 
\bibitem{essen} H. Ess\'en and M.C.N. Fiolhais, \emph{Am. J. Phys.} \textbf{80(2)} (2012) 164. 
\bibitem{fiolhais2} M.C.N. Fiolhais and H. Ess\'en, \emph{Int. J. Theor. Phys.} \textbf{52} (2013) 1701.
\bibitem{halperin} B.I. Halperin, T.C. Lubensky \and S. Ma, \emph{Phys. Rev. Lett.} \textbf{32} (1972) 292.
\bibitem{kleinert} H. Kleinert, Lett. \emph{Nuovo Cimento} \textbf{35} (1982) 405.
\bibitem{kleinert2} H. Kleinert, \emph{Europhys. Lett.}, \textbf{74} (2006) 889.  
\bibitem{kleinert3} H. Kleinert, \emph{Gauge Fields in Condensed Matter, Vol. I}, World Scientific, Singapore (1989)
\bibitem{abrikosov} A.A. Abrikosov, \emph{Zh. Eksp. Teor. Fiz.} \textbf{32} (1957) 1442.
\bibitem{coleman} S. Coleman and E. Weinberg, \emph{Phys. Rev. D} \textbf{7} (1973) 1888.
\bibitem{weinberg} Erick James Weinberg, \emph{Radiative Corrections as the Origin of Spontaneous Symmetry Breaking}, Harvard University, Cambridge, Massachusetts (1973).
\bibitem{parmenter} R.H. Parmenter, \emph{Phys. Rev.} \textbf{166} (1968) 392.
\bibitem{janocko} M.A. Janocko, M. Ashkin and C.K. Jones, \emph{Physics Letters A} \textbf{43} (1973) 345.
\bibitem{guo} Y. Guo \emph{et al.}, \emph{Science Magazine} \textbf{306(5703)} (2004) 1915.
\bibitem{zgirski} M. Zgirski \emph{et al.}, \emph{Nano Letters} \textbf{5(6)} (2005) 1029.
\bibitem{clark} K. Clark \emph{et al.}, \emph{Nature Nanotechnology} \textbf{5} (2010) 261.
\bibitem{graser} S. Graser \emph{et al.}, \emph{Nature Physics} \textbf{6} (2010) 609.
\bibitem{pippard} A.B. Pippard, \emph{Proc. R. Soc. Lond. A} \textbf{203} (1950) 210.
\bibitem{fiolhais3} M.C.N.~Fiolhais and J.L.~Birman, \emph{Europhysics Letters} \textbf{107} (2014) 27001.
\bibitem{fiolhais4} M.C.N.~Fiolhais and J.L.~Birman, \emph{Physics Letters A} \textbf{378} (2014) 2632.
\bibitem{alberto} P.~Alberto, C.~Fiolhais and V.M.S.~Gil, \emph{European Journal of Physics} \textbf{17} (1996) 19.
\bibitem{fiolhaiskleinert} M.C.N.~Fiolhais and H.~Kleinert, \emph{Phys. Lett. A}, \textbf{377} (2013) 2195.
\end{thebibliography}
\end{document}